# Image Processing in Floriculture Using a robotic Mobile Platform

García Torres Juan Mauricio
Fundación Universitaria Agraria De Colombia
Bogotá, Colombia
juang1145@hotmail.com

Caro Prieto Diana Carolina
Fundación Universitaria Agraria De Colombia
Bogotá, Colombia
carodiana3@gmail.com

*Abstract*— Colombia has a privileged geographical location which makes it a cornerstone and equidistant point to all regional markets. The country has a great ecological diversity and it's one of the largest suppliers of flowers for US. Colombian flower companies have made innovations in the marketing process, using methods to reach all conditions for final consumers. Nowadays, Robotic has performed a significant role in agriculture thanks to its the level of accuracy. Examples of this are all developments in computer vision, allowing greater control over production processes such as quality inspection, selection and classification of a product according to market requirements, ensuring high quality standards in production for exporting, optimizing performance times at the production level, among others.

Computer vision is based on image processing, method used for identifying environment objects and making decisions about their behavior. The main tasks of the vision system are detecting, segmenting and recognizing patterns, integrating systems of digital image capture, input-output devices and computer networks.

The image processing in real time using artificial vision systems applied in floriculture is mainly focused on the use of techniques and methods to obtain, process and analyze the information provided by the images of different types of flowers to export. In this case, fulfilling the goal of quality inspection system checks the compliance of a flower with certain characteristics in their appearance, such as size, uniformity of color, brightness, etc.

This article develops a monitoring system for floriculture industries. The system was implemented in a robotic platform. This device has the ability to be programmed in different programming languages. The robot takes the necessary environment information from its camera. The algorithm of the monitoring system was developed with the image processing toolbox on Matlab. The implemented algorithm acquires images through its camera, it performs a pre-processing of the image, noise filter, enhancing of the color and adjusting the dimension in order to increase processing speed. Then, the image is segmented by color and with the binarized version of the image using morphological operations (erosion and dilation), extract relevant features such as centroid, perimeter and area. The data obtained from the image processing helps the robot with the automatic identification of objectives, orientation and move towards them. Also, the results generate a diagnostic quality of each object scanned.

I. INTRODUCTION

The flower industry is undoubtedly one of the most successful businesses in the country; it's the third largest foreign exchange earner after oil and coffee. It has been an important employer generator in Colombia, evidenced over 30 years. The Colombian floriculture has exceeded sales up more than $ 500 million annual [1], which has led Colombia today be the second largest exporter of flowers in the world after The Netherlands [2].

However, this industry has several problems: the lack of oversight on the product selection since the harvest period depending on the features required to export, the use of high residual substances which causes pollution in the setting and early discoloration petals [3]. Other issues mentioned are the control of temperature, humidity and pests, but these depend on the environment where they were developed.

A possible solution to increase the quality of the product exported to other countries consists on applying the technology on the agriculture sector. They have done different studies regarding the application of image processing in agriculture on coffee [4].

Also, another study was done using morphological operations in the coffee with three different methods using the technical of region's growth. The result suggested on the reseach recommends the implementation of systems with discontinuity information, pixel by pixel and region-and contour [5].

Some process have been made using PLC to classify different objects depending on the color and material, using optical and inductive sensors [6].

Several industrial applications made on products like coffee, have also been studied in flowers, focused on their behaviors such as the way they open and close their petals, growth and their adaptation in the environment [7].

In base on the background studied, we propose the use of automation in floriculture, because companies can ensure the quality on the products they sell, using sensors in order to identify the proper color, controllers to program the process and actuators to separate, cut or select the final product. In this article, the researchers are focused on the interpretation of the signals given by the sensors (in this case, the images obtained by the camera) and the programming of the robot for monitoring and selecting the right flower.

## II. METHODS

### A. Object to analyze

The Carnation (Dianthus caryophyllus) is one of the most important flowers in Colombian floriculture companies and by its beauty, is one of the most exported flowers in the country.

The features of a carnation are:

- The stem must be strong and straight with the sheets clean and wide.
- The pedicel must be straight, keeping the flower firm.
- The petal color is net and the petals are not banded.
- Must have a good vase life. The size of the flower depends on its temperature.

The colors observed in carnation petals are red, pink, white, salmon, yellow and bicolor. [8].

### B. Mobile Robotics Platform

The robot used has different actuators which allow its operation with respect to the environment where it's located, such as motors, distance sensors and camera. It allows the data acquisition using input devices and performing a respective image processing depending on the software used. In this case we used Matlab platform.

Matlab is mathematical software that allows the execution of programs depending on the needs that the user has.

### C. Image Processing

The image processing is used for making the image treatment and it must show some important details for its selection.

The image treatment can be performed in real time in order to observe the results in terms of industrial application even if the process is running.

With the image processing, companies can recognize an object with certain characteristics in order to make a selection and verification of the picture taken (possibly in real time).

### D. Procedure

The program developed does the pre-processing and image processing, determining some characteristics about color, size and texture. Applying a pre-processing of the image allows getting an efficient development of work since the tests phase. It will be constantly changing according to environment conditions and with some variables such as the brightness [9], for that reason, we used a solid image.

The steps followed for the project were:

- Acquisition of the image.
- Pre-processing and image processing
- Image classification based on the parameters selected.
- Binarization.
- Application of morphological operations

**Acquisition of the image**

The project is focused on the carnations, but during the experiment it was taken as reference a picture of a single flower in order to have a pattern for the characterization and implementation avoiding continual changes in the details of a flower naturally.

The mobile platform allows the acquisition of images using its own camera with the following features: VGA CMOS, 640x480 resolution video at 15fps, 1024x768 resolution images, format still image capture to BMP and JPG, color depth 24-bit true color.

In concepts of artificial vision, the image segmentation identifies individually each element in the environment [10].

In this project, the segmentation was done just to identify the flower with specific features, without consider other objects. The best way to choose of a specific pixel area is by obtaining the component of the RGB space for the color selected in the same image, in this case, the red.

**Characterization**

In the characterization, physical properties as color, shape and size must be considered. To obtain a stationary contour, it has to be applied an average low pass filter with a circular mask with a radius of 5 px or a rectangular mask of 5x5, with the objective to delete noise that it can affect the selection of color. After, It can make a segmentation with RGB characteristics where color red can be chosen, controlling the image intensity and getting an effective binarization.

Binarization is the process where images can be transformed into binary numbers, being white with a value of one and black with a value of zero with respect to the color of the pixel selected previously. This process allows obtaining an exact color of the product desired. One of the advantages of this process consists on identifying the error in the product, such as Anthractnose, Botrytis gray mold, Black spots, rust, canker [11].

With operations like erosion and dilation (morphologic operations), irrelevant details in the selection of color can be deleted. Further, it expands the area of the tone, without deforming the figure.

The image chosen to the experiment has 399x515 of size, in PNG format and red color (See Fig. 1).

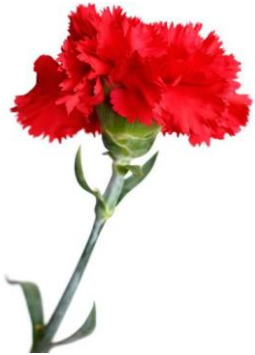

*fig. 1. Original image [12]*

### III. RESULTS

Representative pixel picked in a RGB space (See Fig 2).

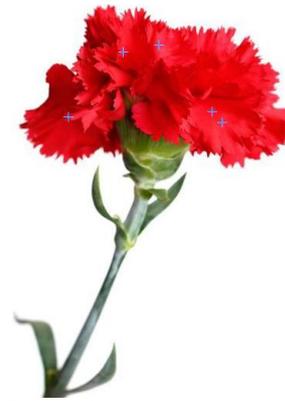

*fig.2. Image with representative pixels [13].*

A part of the image is selected in order to facilitate the processing of the image (see Fig 3).

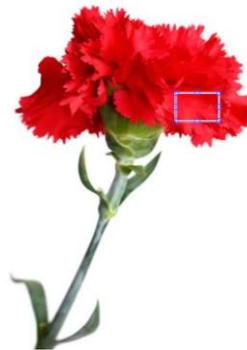

*fig.3.Segmentation indicated [14].*

Segment selected, this image visualizes the different samples of classification with the variable of color, making a range of the color according to its tone, in this case, red (See Fig. 4)

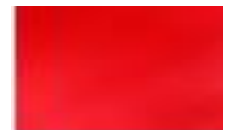

*fig.4.Segment selected.*

**Morphologic operations**

With the characteristics obtained, the binarization can be performed, identifying the red color on the image (See Fig. 5).

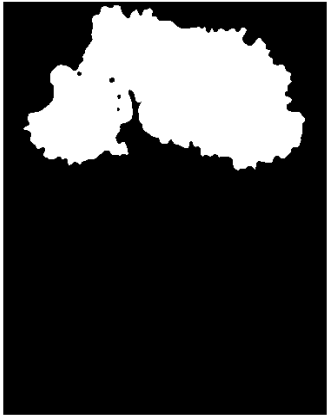

fig.5.Image binarized [15].

**Communication with the robot**

When the image is binarized in Matlab, the controller of robot proceeds to perform the program code and after, the program generates a log, informing the operation about the characteristics of the flower and its status.

The communication with the robot is made with the WLAN configuration using a network address according to the IEEE 802.11g and 802.11b standards; it allows establishing a safe network with WEP password and WPA-PSK function.

IV. CONCLUSIONS

This article shows a system classifier of flowers using sensors in order to obtain data of physical characteristics of the flower according to its color. Image processing algorithms where implemented under the standards of the platform Matlab.

With the image processing, several applications can be performed in the floriculture sector, giving profits to those companies, increasing the velocity of the process and the percentage of quality on the product.

Automation can classify the flower with variables like color, obtaining a better classification of this product for its exportation.